\documentclass{article}
\usepackage{spconf,amsmath,graphicx}

\usepackage{amsmath}
\usepackage{subcaption}
\usepackage{comment}
\usepackage[capitalise]{cleveref}

\newcommand{\norm}[1]{\left\lVert#1\right\rVert}
\newcommand{\abs}[1]{\left\lvert#1\right\rvert}

\title{Cortical Features for Defense Against Adversarial Audio Attacks}
\name{Ilya Kavalerov$^1$, Ruijie Zheng$^1$, Wojciech Czaja$^1$, Rama Chellappa$^2$\thanks{Ilya Kavalerov and Rama Chellappa acknowledge support of the MURI from the Army Research Office under the Grant No. W911NF-17-1-0304. This is part of the collaboration between US DOD, UK MOD and UK Engineering and Physical Research Council (EPSRC) under the Multidisciplinary University Research Initiative. Wojciech Czaja acknowledges support of the NSF DMS grant \# 1738003.}}
\address{$^1$University of Maryland, College Park MD\\ $^2$Johns Hopkins University, Baltimore MD.}
\begin{document}
%
\maketitle
\begin{abstract}
We propose using a computational model of the auditory cortex as a defense against adversarial attacks on audio.
We apply several white-box iterative optimization-based adversarial attacks to an implementation of Amazon Alexa's HW network, and a modified version of this network with an integrated cortical representation, and show that the cortical features help defend against universal adversarial examples.
At the same level of distortion, the adversarial noises found for the cortical network are always less effective for universal audio attacks.
\end{abstract}
\section{Introduction}
\label{sec:intro}
As voice assistant systems like Amazon Alexa, Google Assistant, Apple Siri and Microsoft Cortana become more ubiquitous and integrated into modern life, so does the risk of antagonists taking control of devices that we depend on.
Adversarial Attacks on Audio \cite{fgsm} are one way that a voice assistant could be subverted to send an unwanted message, or bank transfer, or unlock a home.
All the mentioned voice assistants rely on wake-word detection to initiate their automatic speech recognition (ASR) systems that give control over their functions.
The wake word detection system is both a voice assistant's first line of defense against adversarial attacks and its least complex feature.
%
\begin{figure}[t!]
	\centering
		\centering
		\includegraphics[width=1\linewidth]{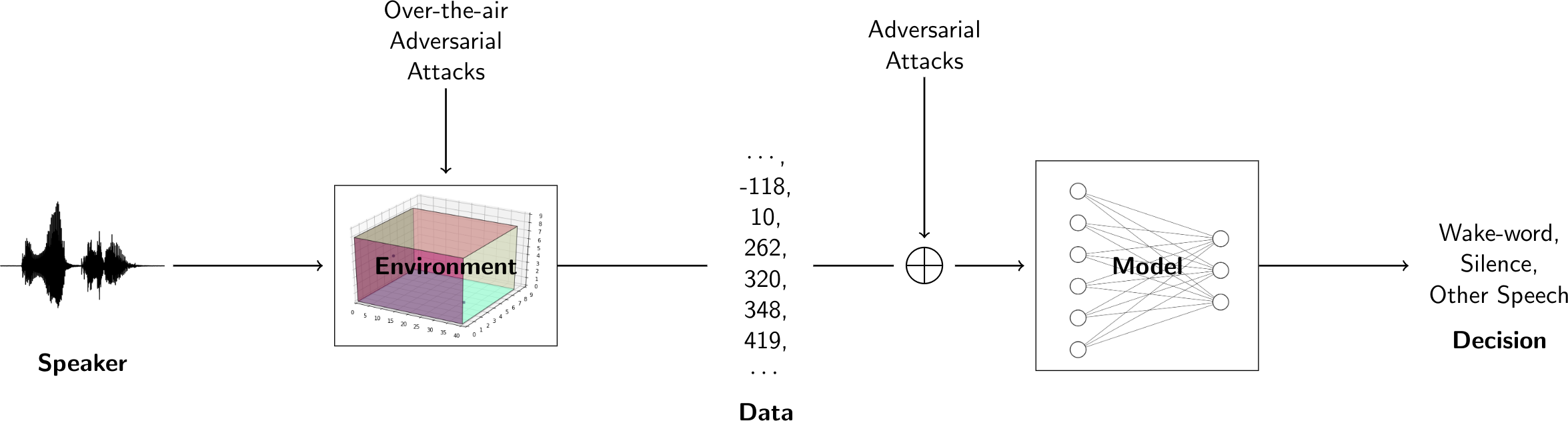}
	\caption{The flow of information for a voice assistant wake-word detector and two attack vectors. The speaker generates audio for their intended command. The environment is an acoustic environment containing the speaker, other sources of audio like background and other speakers, distortions like reverberations, and the voice assistant's microphone and audio hardware including the ADC. The data (assumed to be 16 bit integer and 16kHz) is then available to a model which ultimately makes the decision whether or not to "wake up" its general ASR capability.}
	\label{fig:adv_channel}
\end{figure}

A number of recent works have achieved adversarial examples for automatic speech recognition.
The Carlini-Wagner attack \cite{cw} can produce an adversarial waveform that is 99.9\% similar to the original, but produces any other intended sentence by the ASR.
Adversarial examples have since developed and improved in several ways.
An important attribute for practical adversarial audio is that it be causal.
Often an adversarial example is a function of the complete original audio that it is meant to modify, so thus original audio from hundreds of milliseconds in the future must be accessed to play an adversarial sound in the present.
One solution is {\it universal adversarial examples} \cite{univadv}, examples that can behave adversarially regardless of what original audio they are superimposed on.
Another important attribute for practical adversarial audio is that it be robust, that is the adversarial example remains effective despite some level of distortions.
This goal is not completely orthogonal to universal adversarial examples, since examples cannot be universal without also being invariant to translation in time, and when superimposed with some other audio.
The ultimate test of robustness is over-the-air attacks, where as shown in \cref{fig:adv_channel} the adversarial example is introduced to the environment generating the data for the model rather than to the data directly.

Robust adversarial examples have been demonstrated to exist for simulated environment ASR systems \cite{qin_2019}, and more recently robust and universal examples have been generated for ASR systems, and even commercial wake word systems \cite{adv_music}. 
%
\begin{figure*}[t!]
	\begin{minipage}[b]{.7\linewidth}
		\centering
		\centerline{\includegraphics[width=1\linewidth]{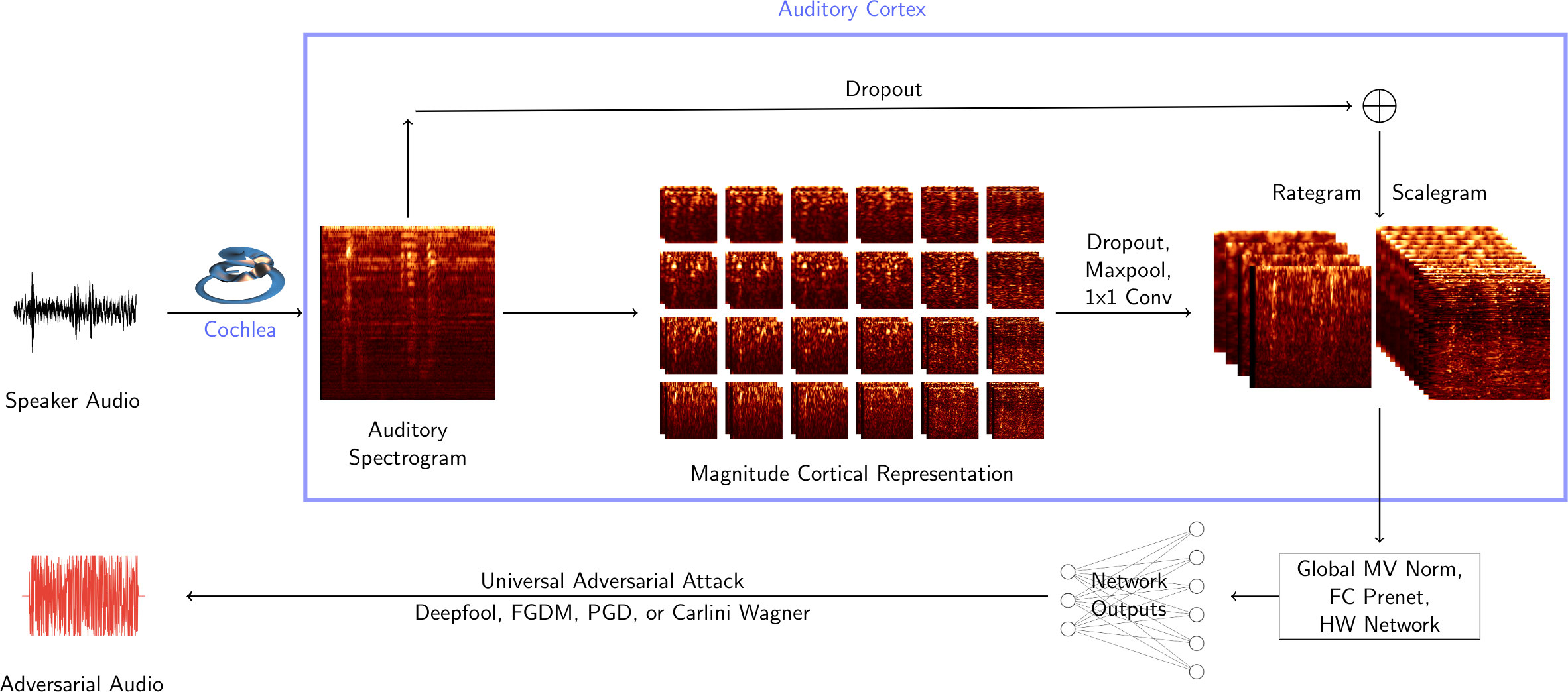}}
		\centerline{(a) Our proposed cortical network}\medskip
	\end{minipage}
	\hfill
	\begin{minipage}[b]{0.29\linewidth}
		\centering
		\centerline{\includegraphics[width=1\linewidth]{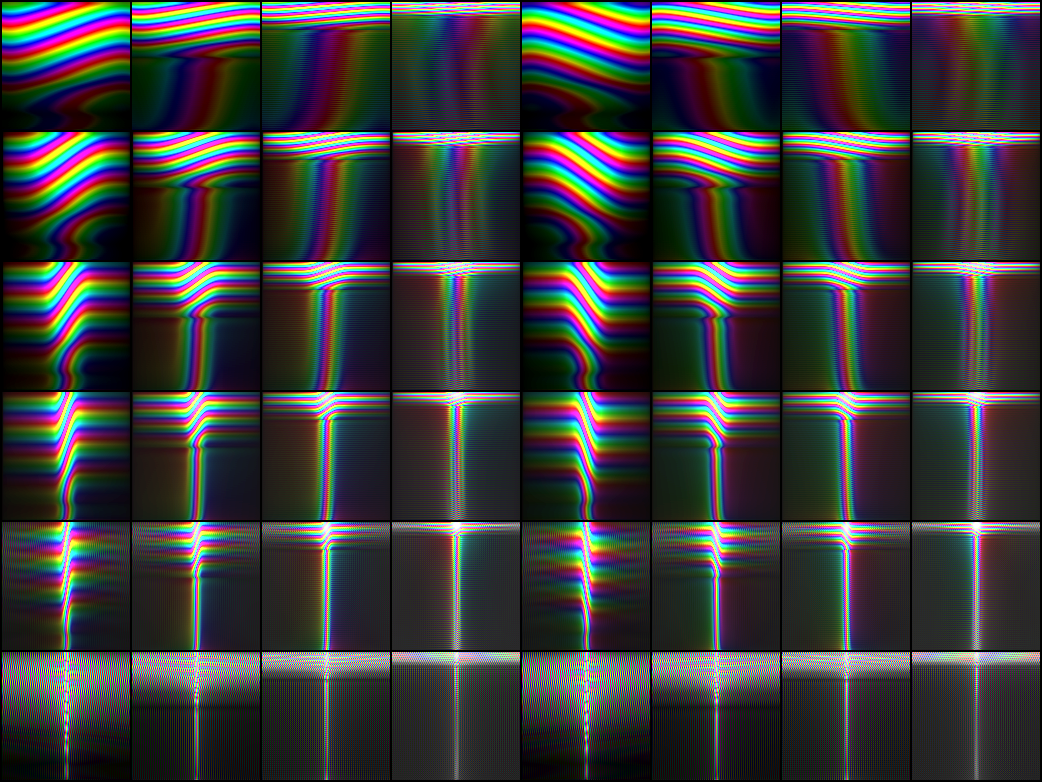}}
		\vspace{.75cm}
		\centerline{(b) Complex colorized STRFs}\medskip
	\end{minipage}
	
	\caption{Our proposed cortical network (a), meant to emulate the first two stages of audio processing in the brain (labeled in light blue). Integrating this features extraction directly into the network gives us two advantages over pre-processing: 1) The feature representation can be enhanced with learning (dropout with 1x1 convolutions) 2) We can backpropagate through these layers are find adversarial audio directly without any lossy transformations (such as Griffin-Lim). The complex STRFs of (a) are shown in (b).}
	\label{fig:cortical_net}
\end{figure*}
Across many languages speech has a pattern of spectrotemporal modulations that differ from most environmental noise and distortions \cite{ding_2017}, and the brain has developed a hierarchical neural system that uniquely responds to these signals \cite{lyon_book}.
These specialized neuronal structures are in the auditory cortex, and for example are largely unresponsive to pure tones \cite{lyon_book}.
It has been demonstrated in animals that the auditory cortex encodes vocalization features while suppressing or being invariant to the noise features, and invasive experiments in humans also have shown that auditory cortex neurons selectively suppress of acoustic features of noise in their neural responses \cite{khalighinejad_2019}. 
%

This characteristic of speech-noise separability found the auditory cortex is exactly the same characteristic we would like in artificial intelligence voice assistants.
For example, if a wake-word system had an intermediate representation that was tuned to speech the same way our brains are, then it would be necessary to perceptively modify speech to change that representation and ultimately the output of the wake-word model, making imperceptible adversarial examples impossible.
Indeed the aforementioned biological findings have inspired a computational model of the mammalian auditory cortex to create such a representation \cite{chi_2005,nsl_doc}.

Our aim in this work is to incorporate the advantages of this cortical representation into a voice assistant and show the efficacy of this representation for defending against adversarial audio attacks.
%
%
%
%
\section{Background}
\subsection{Cortical transform}
The details of human hearing have been studied for centuries. In this section we highlight some of the consensus of the last few decades of modeling for the first stages of how the brain makes sense of speech. 
In the early stage, acoustic signals are transformed into 2 dimensional spectrograms in the cochlea, which acts as a filter bank, and filtered by the hair cell and lateral inhibitory network \cite{lyon_book,mesgarani_2011}.
These lower auditory areas are tonotopically organized: each cochlear location is associated with a best frequency it responds to, with time as the other axis this representation can be interpreted as an auditory image \cite{lyon_book}.
As signals move out to the auditory cortex (A1), what was essentially 1 dimensional processing in the cochlea becomes 2 dimensional in A1 \cite{nsl_doc}.
Invasive experiments have measured the responses of ferrets' A1 neurons and characterized them with corresponding spectro-temporal receptive fields (STRFs) \cite{mesgarani_2008}.
An STRF of a neuron indicates the frequencies and time lags that correlate with an increased response of that neuron; these two dimensions are often referred to as scale (spectral) and rate (temporal).
%
A computational model for these receptive fields can be estimated from the measurements of these neurons responses \cite{shamma_95,nsl_doc}.
On the scale axis the transfer function of these neurons are well approximated by the second derivative of a Gaussian function \cite{nsl_doc}.
Neurons are tuned to be responsive to a variety of rates and scales \cite{mesgarani_2008},
for scale features we parameterize the approximation of the neural response as:
\begin{equation}
R_{2,\psi}(y) = (\Omega_2 y / \psi)^2)\exp{(1-(\Omega_2 y / \psi)^2)} 
\end{equation}
\noindent
where $R_{2,\psi}(y) = \mathcal{F}\{ r_{2,\psi}(f) \}$ \cite{nsl_doc}.
On the rate axis, the response is modeled by a function with a central excitatory band surrounded by inhibitory side bands, this is approximated with:
\begin{equation}
r_{1,\omega}(t) = (t/\Omega_1\omega)^2\omega\exp{(-\alpha t / \Omega\omega)}\sin (2\pi t \ \Omega_1\omega)
\end{equation}
\noindent
where $\alpha=3.5$ was chosen empirically, as detailed in \cite{nsl_doc}.
Finally the 2D STRF filters are
\begin{equation}
f(\omega,\psi,\phi) = \begin{cases}
r_{2,\psi} \otimes r_{1,\omega} \ , & \phi = 1 \\
r_{2,\psi} \otimes \overline{r_{1,\omega}} \ , & \phi = -1
\end{cases}
\label{eq:strfs}
\end{equation}
\noindent
When phase is $-1$ the rate filter is conjugated.
These filters with $\psi\in\{.25, .5, 1, 2, 4, 8\}$ (cyc/oct) and $\omega\in \{4, 8, 16, 32\}$ (Hz) are displayed in \cref{fig:cortical_net}.
%
%
%
Taking these biophysical observations, we can build an audio processing pipeline that transforms sound into a two dimensional auditory spectrogram, and then applies the STRFs in \cref{eq:strfs} to form a higher dimensional {\it cortical representation} from the magnitude response. This abstraction captures important physiological observations: selectivity to combined spectral and temporal features, and with proper choice of $\mathcal{W}$ the temporal dynamics of phase locking which decreases to less than 30 Hz in the cortex \cite{chi_2005} (something only possible with large filters).
This pipeline can be seen in \cref{fig:cortical_net}.
The cortical transform of audio is a high dimensional tensor with dimensions (time $\times$ frequency $\times$ rate $\times$ scale).
This can be reduced to two 3D tensors by reducing/max pooling across the rate ({\it scalegram}) or scale ({\it rategram}) \cite{mesgarani_2011}.
%
Such a pre-processing feature extraction pipeline has previously been applied successfully to speaker detection \cite{mesgarani_2006} and phoneme classification \cite{mesgarani_2011}.
The cortical representation captures the distinctive dynamics of phonemes, which are easily visually discernible in rategrams and scalegrams \cite{mesgarani_2011}.
The magnitude cortical response carries enough information to reconstruct the spectrogram that generated it \cite{chi_2005}.
Because of its ability to capture speech specific features, inverting the magnitude cortical response has been applied to speech enhancement since it has high fidelity in the reconstruction of speech and low fidelity in the reconstruction of background noise \cite{mesgarani_2007}.
%
%
%
\subsection{Adversarial attack methods}
We attack wake-word networks $f: x \mapsto y\in \{-1,0,1\}$, which predict other-speech, no speech, wake-word, at each time step. 
In the targeted adversarial attack threat model, the goal is to find a small adversarial noise $\delta$ s.t. $f(x+\delta)=y'\neq f(x)$, where $y'$ is a target label.
That is we minimize $l(f(x+\delta), y')$ s.t. $\norm{\delta}_\infty \leq \epsilon$, where $l$ is cross-entropy.
To solve this minimization problem, the Fast Gradient Sign Method (FGSM) finds the direction of the sign of the gradient and takes a corresponding step in this direction:
$x = x - \epsilon \ \mathrm{sign}(\nabla l(f(x),y'))$
\cite{fgsm}.
More advanced adversarial example generation algorithms have adopted an iterative scheme to solve the optimization task \cite{pgd,deepfool,cw,univadv}, including the projected gradient descent (PGD) attack, the DeepFool attack, as well as Carlini-Wagner (CW) attack. The DeepFool attack iteratively makes a linear approximation of the decision boundary and then takes a step normal to the tangent plane \cite{deepfool}. The CW attack solves a single optimization algorithm $l(f(x+\delta), y')+c\|\delta\|$, where the loss function $l$ has been carefully chosen \cite{cw}. The PGD attack projects the perturbed input to the constrained $\epsilon$-ball to enforce the $\ell_\infty$ constraint. In our paper, we consider these three attack algorithms as "strong" attacks and the goal is to show the efficacy of the cortical network to defend against these adversarial attacks.
\subsection{Wake-word detection}
A variety of networks have been published by team members at Amazon, and the time-delayed bottleneck highway network (TDB-HW) is the latest in the literature \cite{amazon_HW}. 
We choose this network as a baseline because it is a larger and more powerful network than some others in the literature \cite{tinyml_from}.
It consists of a feature extractor stage and a classifier stage.
Before the feature extraction stage, auditory spectrogram (a log-mel filter bank spectrogram, abbreviated LFBE) features are fed into a two layer FC prenet with dropout. 
The feature extraction stage of TDB-HW consists of four highway blocks.
Then, a bottleneck with 20 left (200ms) and 10 right (100ms) contexts is applied.
Finally, six highway blocks and one FC layer are applied to provide a classification.
This is the network we refer to as a "baseline".


\begin{figure}[t!]
	
	\begin{minipage}[b]{.24\linewidth}
		\centering
		\centerline{\includegraphics[width=1\linewidth]{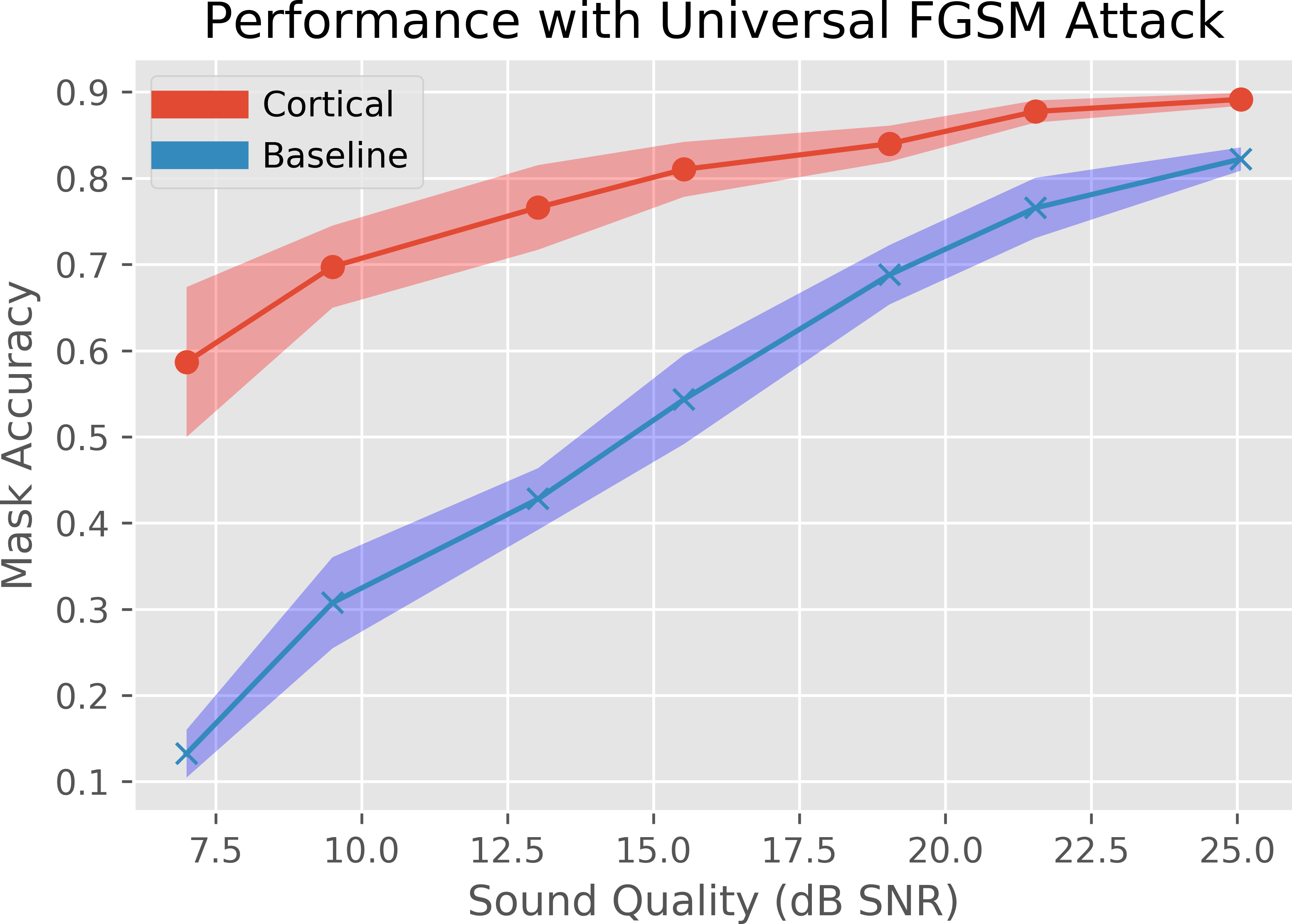}}
	\end{minipage}
	\hfill
	\begin{minipage}[b]{0.24\linewidth}
		\centering
		\centerline{\includegraphics[width=1\linewidth]{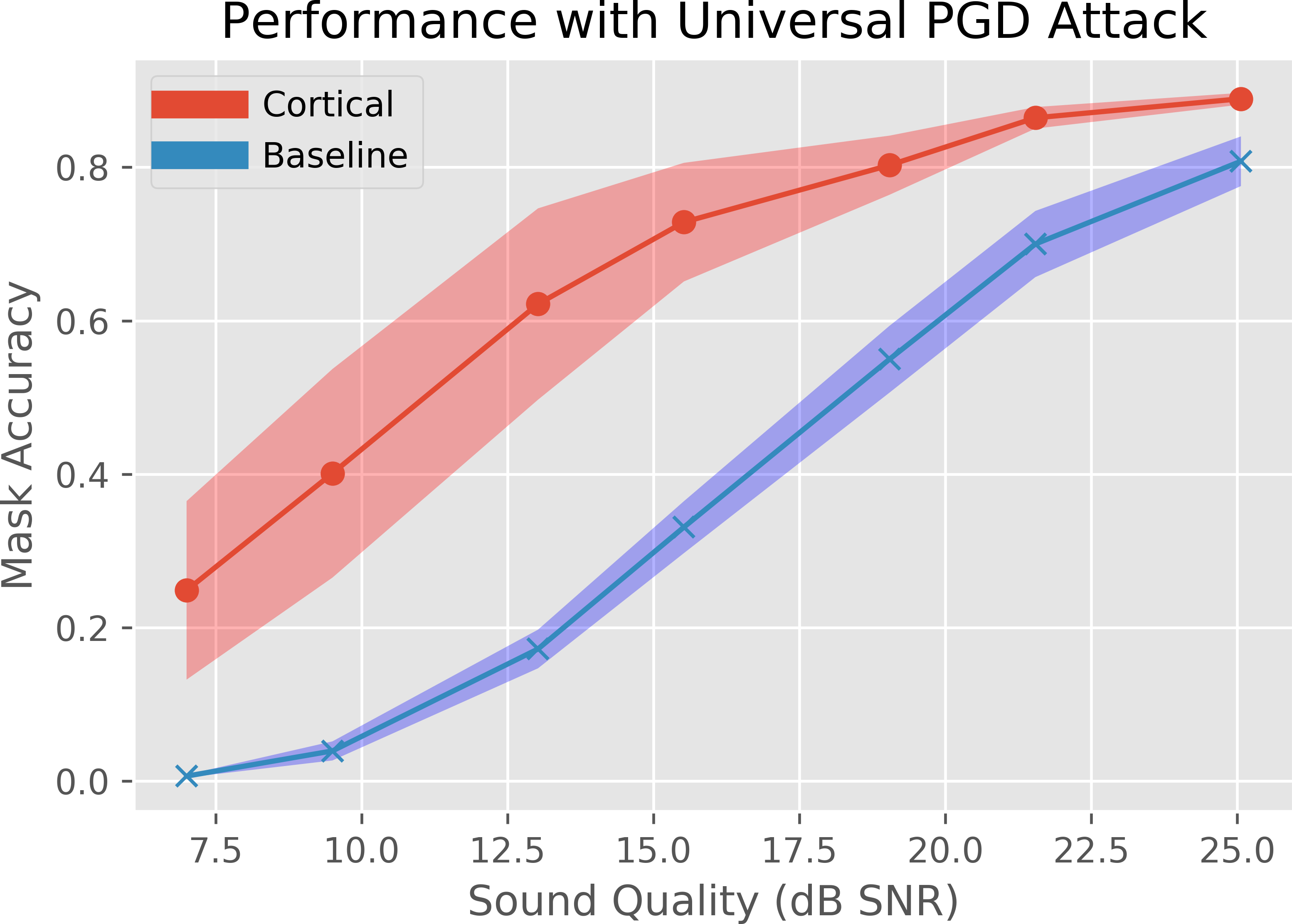}}
	\end{minipage}
	\hfill
	\begin{minipage}[b]{0.24\linewidth}
	\centering
	\centerline{\includegraphics[width=1\linewidth]{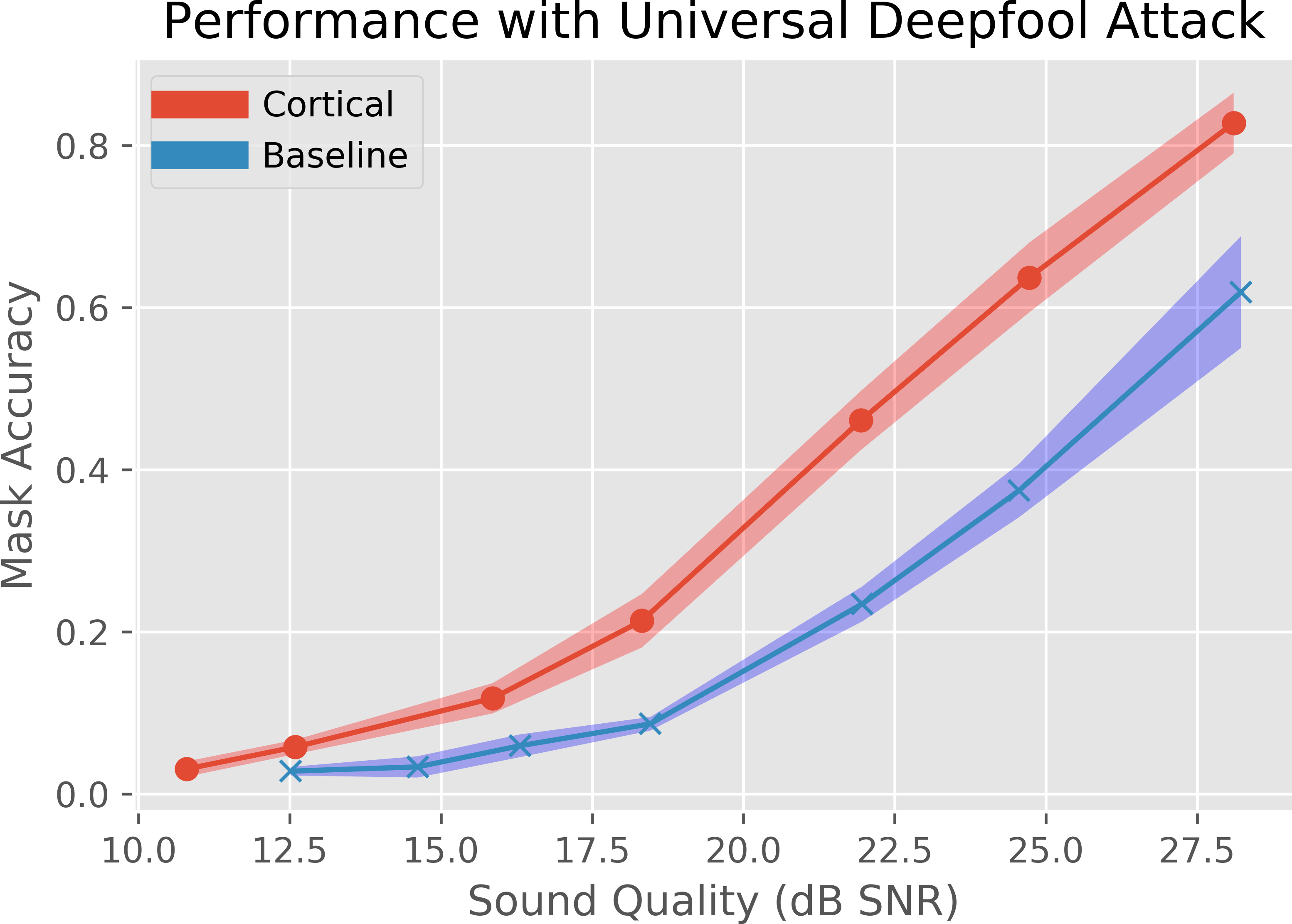}}
	\end{minipage}
	\hfill
	\begin{minipage}[b]{0.24\linewidth}
	\centering
	\centerline{\includegraphics[width=1\linewidth]{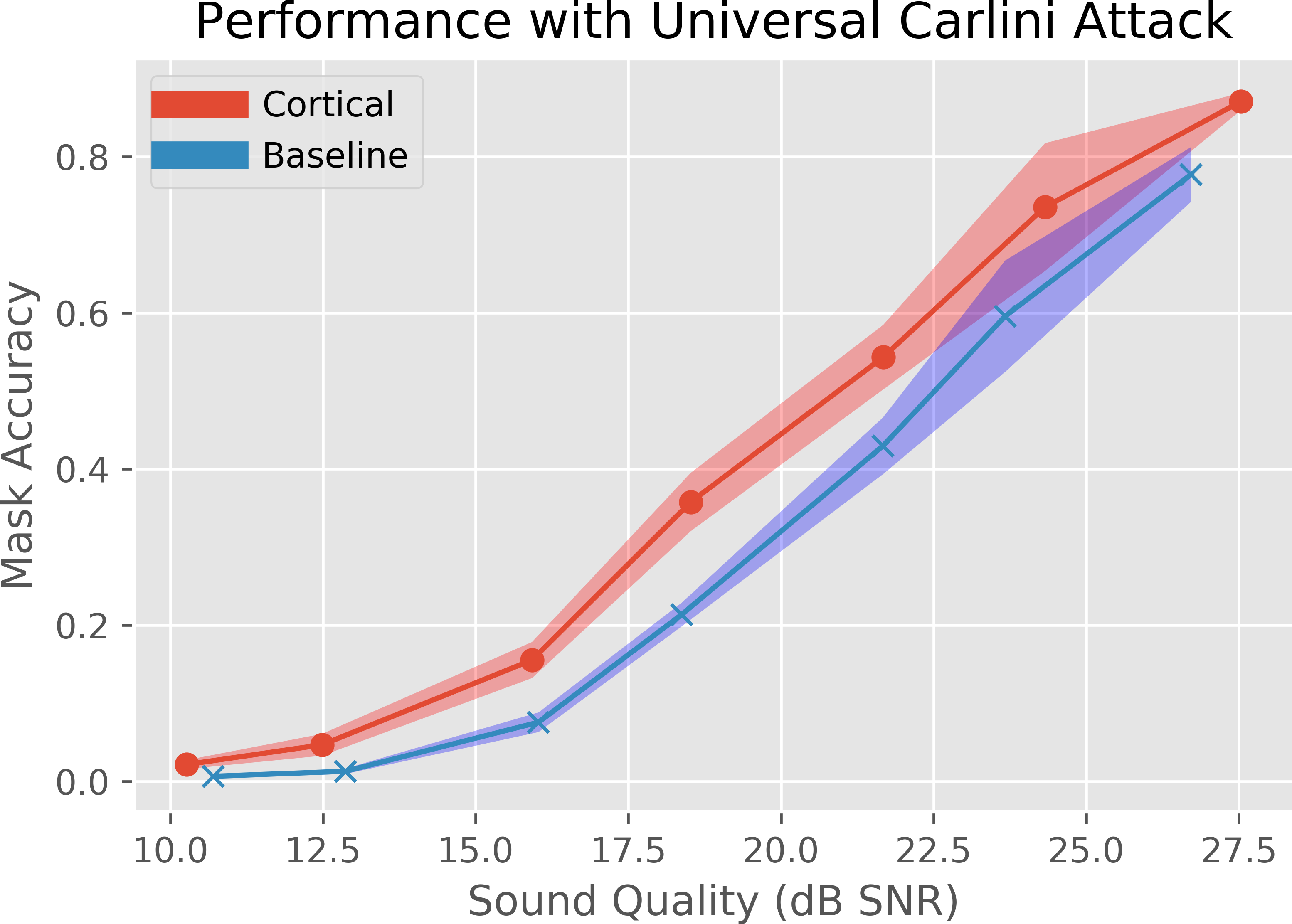}}
	\end{minipage}
	\caption{
		Three baseline and three cortical networks are attacked in four trials each with various SOTA attacks from IBM's ART toolbox \cite{ibmart}.
	Each point in each figure corresponds to the hyperparameters for a single type of attack constrained to an $\ell_\infty$ ball. 
	Each point is the mean of 12 trials, showing the mean test mask accuracy for the best (minimum accuracy) adversarial noise found during the attack trial on a different attack training set.  1 standard deviation is shaded.}
	\label{fig:defense1}
\end{figure}

\begin{figure}[t!]
	
	\begin{minipage}[b]{.32\linewidth}
		\centering
		\centerline{\includegraphics[width=1\linewidth]{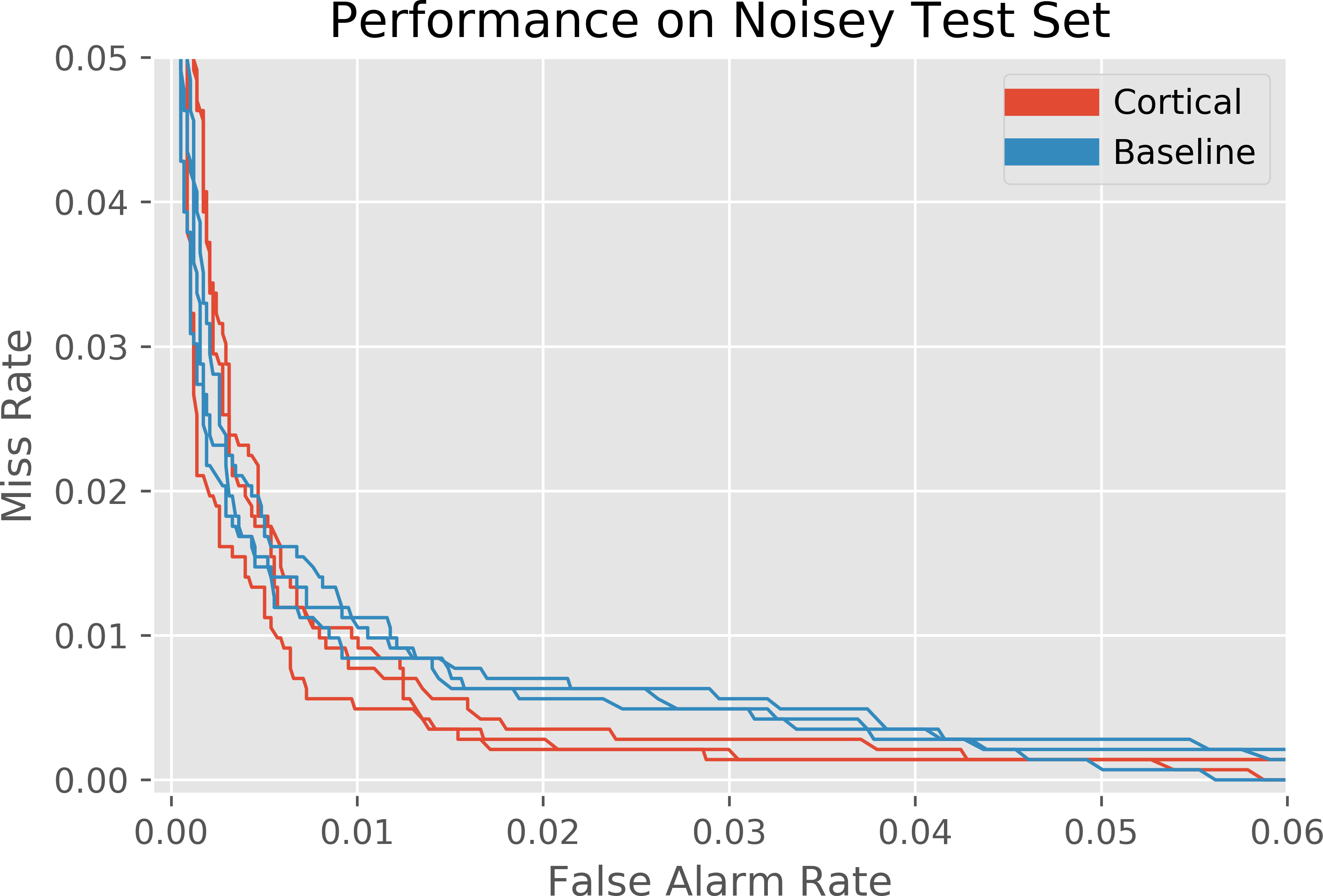}}
	\end{minipage}
	\hfill
	\begin{minipage}[b]{0.32\linewidth}
		\centering
		\centerline{\includegraphics[width=1\linewidth]{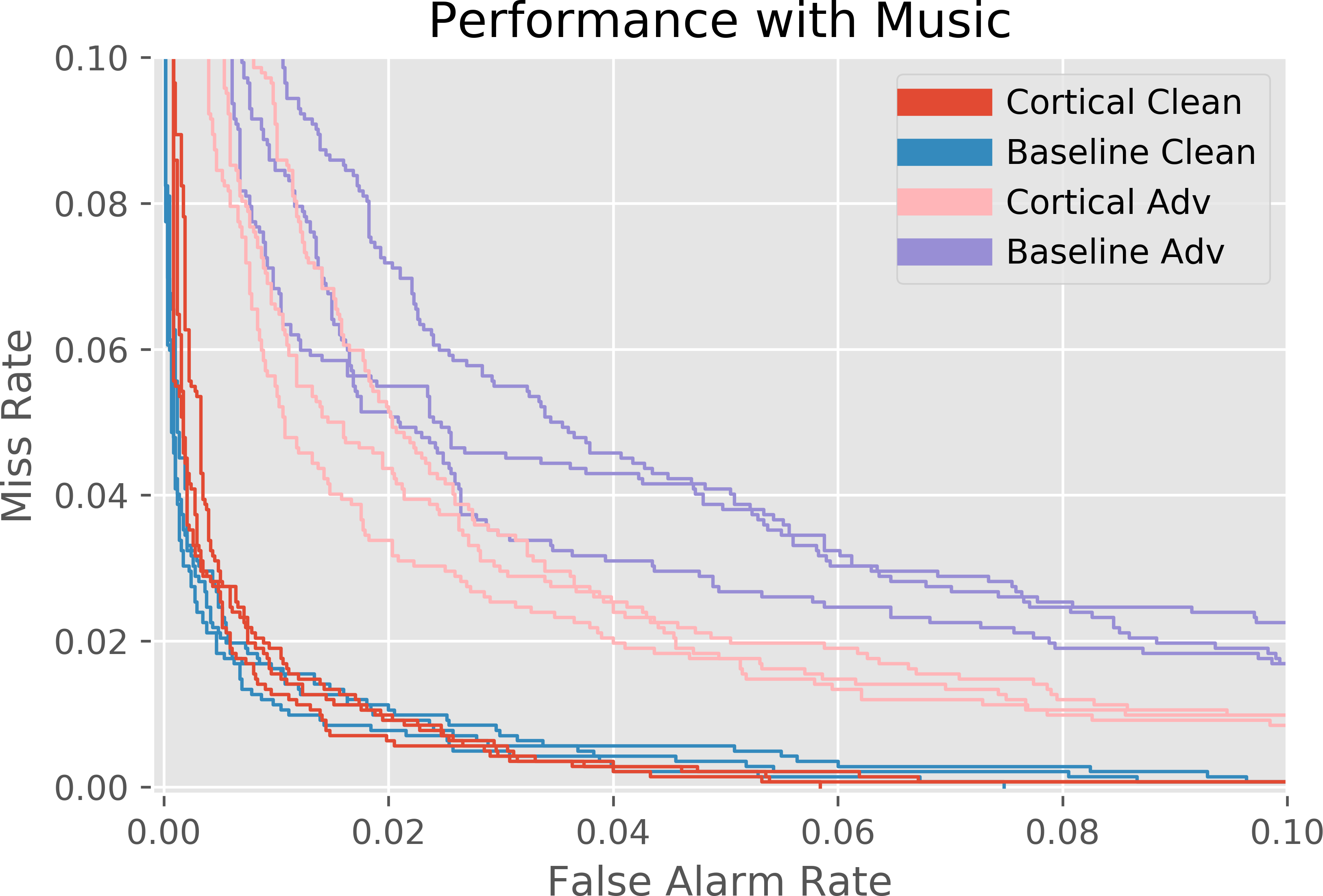}}
	\end{minipage}
	\hfill
	\begin{minipage}[b]{0.32\linewidth}
		\centering
		\centerline{\includegraphics[width=1\linewidth]{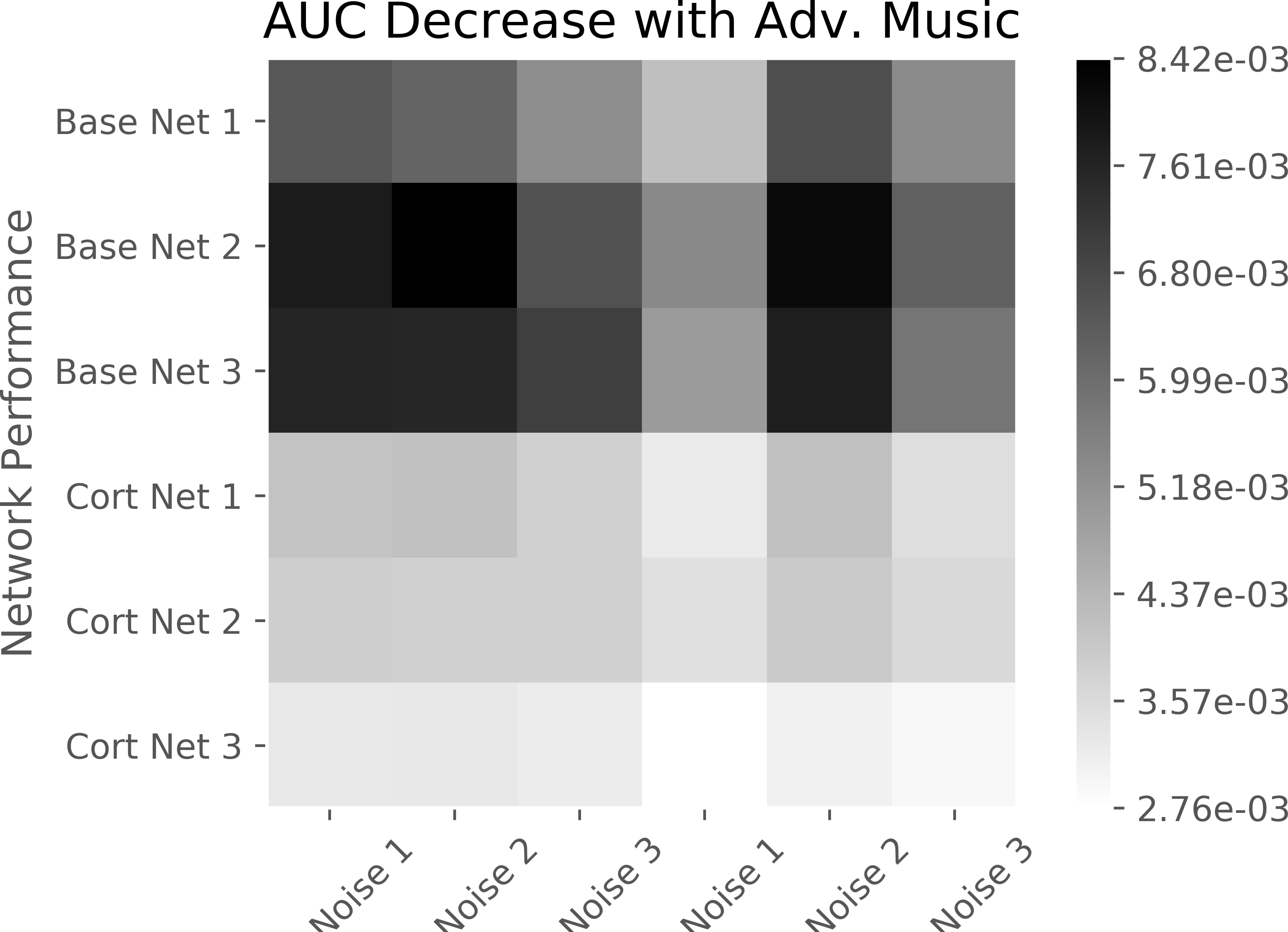}}
	\end{minipage}

	\caption{False alarm and Miss rates of the networks that we compare. One line is plotted for each of three training trials for each of the two architectures. Left test set performance, Middle performance with adversarial music \cite{adv_music}, and Right shows the AUC of the baseline networks decreases the most with adversarial music.}
	\label{fig:perf}
\end{figure}

\section{Proposed Architecture}

For our cortical network, the input LFBE features are convolved with 48 size 32x32 complex STRF filters from \cref{eq:strfs}. These 48 filters are shown in \cref{fig:cortical_net}, and have parameters which mimic the response measured in auditory cortex neurons \cite{nsl_doc}. After applying an absolute value and dropout, max-pooling is applied along the rate and scale axes of the features to yield a scalegram and a rategram. These are concatenated, a 1x1 convolution is applied, and the original spectrogram with substantial dropout (0.9) is added to this 3D tensor. A small prenet is applied before the TDB-HW network is applied. The residual addition is motivated by the loss of 2D high frequency features by the STRFs, and a way to recover some of that information.
The STRFs are not learned during training, so the total learnable parameters are similar between the two models.

\section{Experimental Results and Discussion}
\subsection{Methods}

We created several monaural 15 hour training datasets and 3 hour validation and test sets.
We create our train/val/test splits from different speakers of wake-words from the Kaggle Alexa dataset and speakers in the Librispeech dataset. 
For additional background noise we use the DEMAND dataset kitchen, living room, laundry room, hallway, office, cafeteria, cafe environments.
We extract words from our positive and negative datasets, pitch shift and time stretch these words, and mix them in a room, randomizing all these variables according to reasonable distributions. 

We evaluate the merit of the cortical network as a defense to adversarial examples on Universal FGSM, PGD, Deepfool, and Carlini-Wagner attacks \cite{fgsm,pgd,deepfool,cw,univadv}. We use the standard implementation of these in the IBM Adversarial Robustness Toolbox (ART) \cite{ibmart}.
The FGSM, Deepfool, and Carlini-Wagner attacks were performed for 4000 iterations and evaluated every 400, PGD was performed for 250 examples for 100 iterations each, evaluated every 25 examples.
For each trial, the minimum accuracy noise was recorded on a test set, and the average minimum accuracy of 12 trials over the three networks is show in \cref{fig:defense1}.

\begin{figure}[t!]
	
	\begin{minipage}[b]{.24\linewidth}
		\centering
		\centerline{\includegraphics[width=1\linewidth]{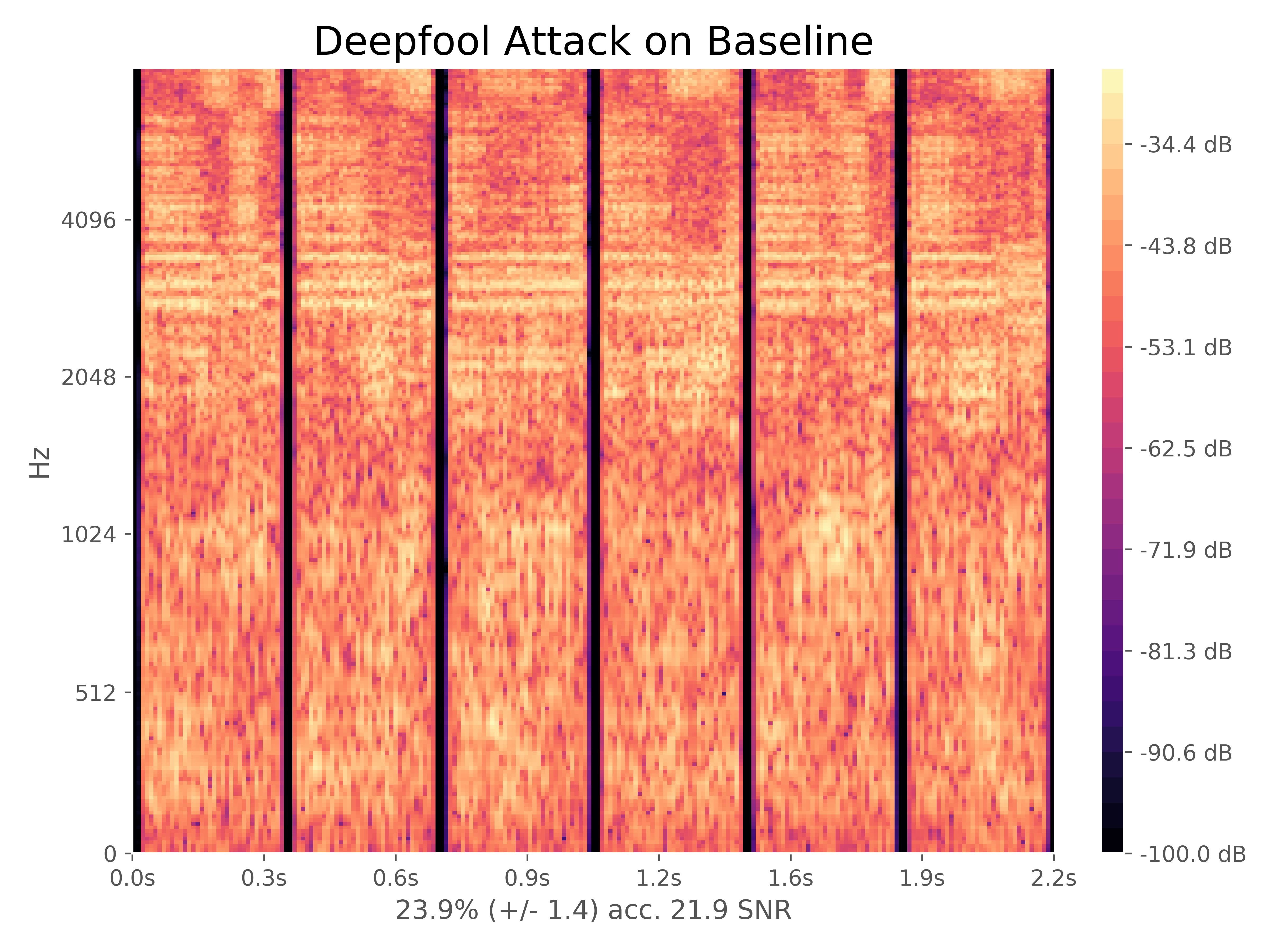}}
	\end{minipage}
	\hfill
	\begin{minipage}[b]{0.24\linewidth}
		\centering
		\centerline{\includegraphics[width=1\linewidth]{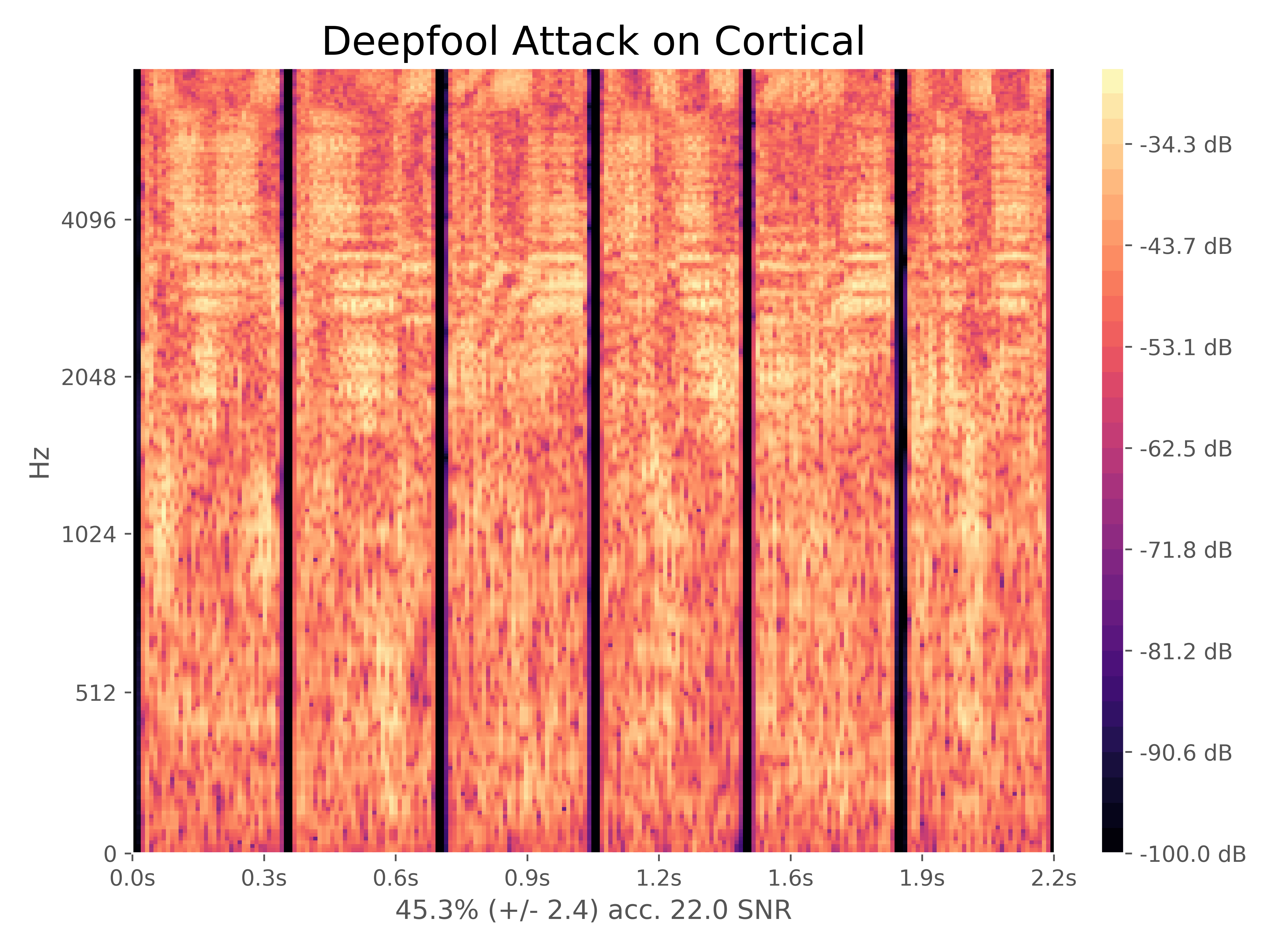}}
	\end{minipage}
	\hfill
	\begin{minipage}[b]{0.24\linewidth}
		\centering
		\centerline{\includegraphics[width=1\linewidth]{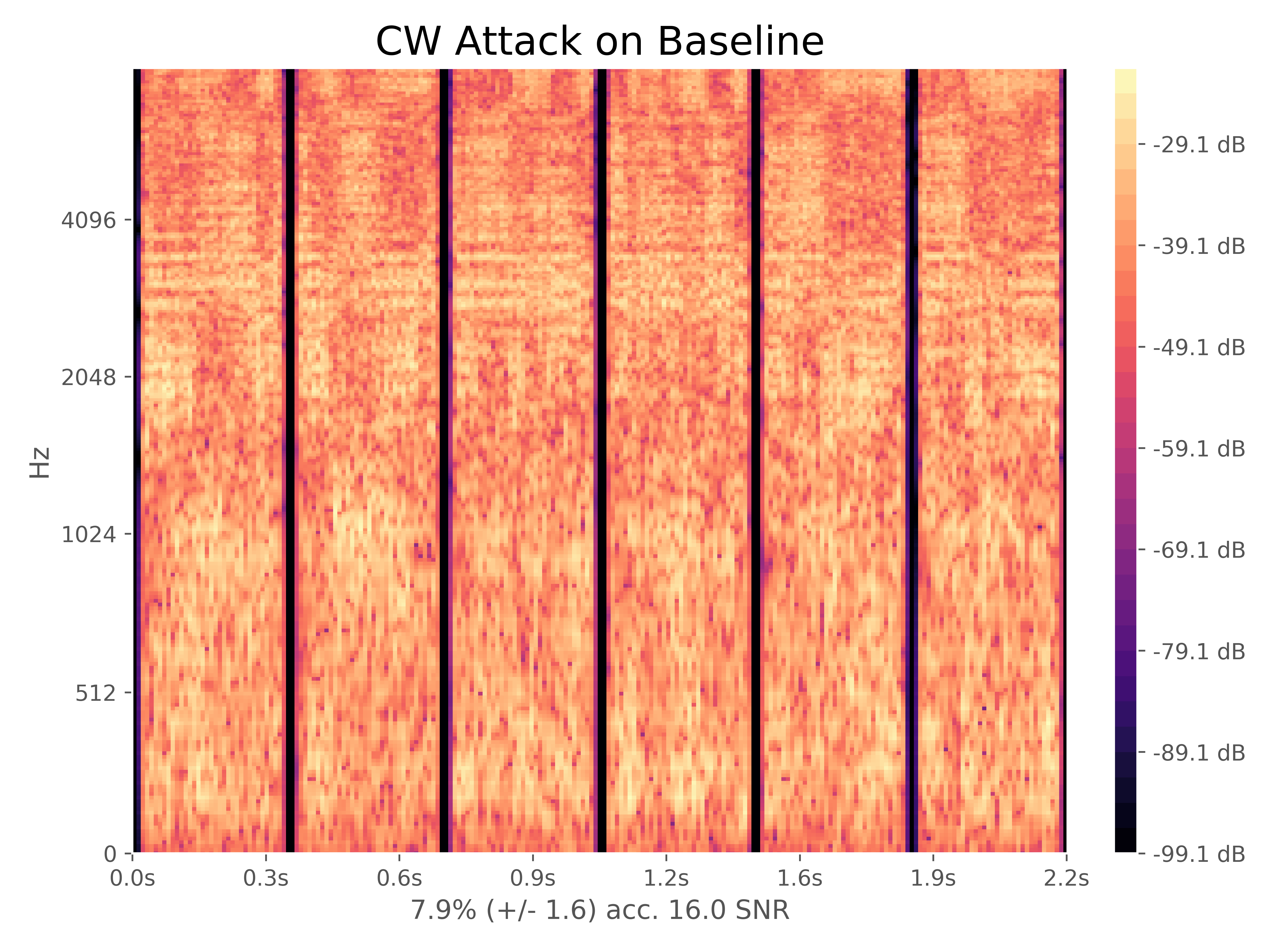}}
	\end{minipage}
	\hfill
	\begin{minipage}[b]{0.24\linewidth}
		\centering
		\centerline{\includegraphics[width=1\linewidth]{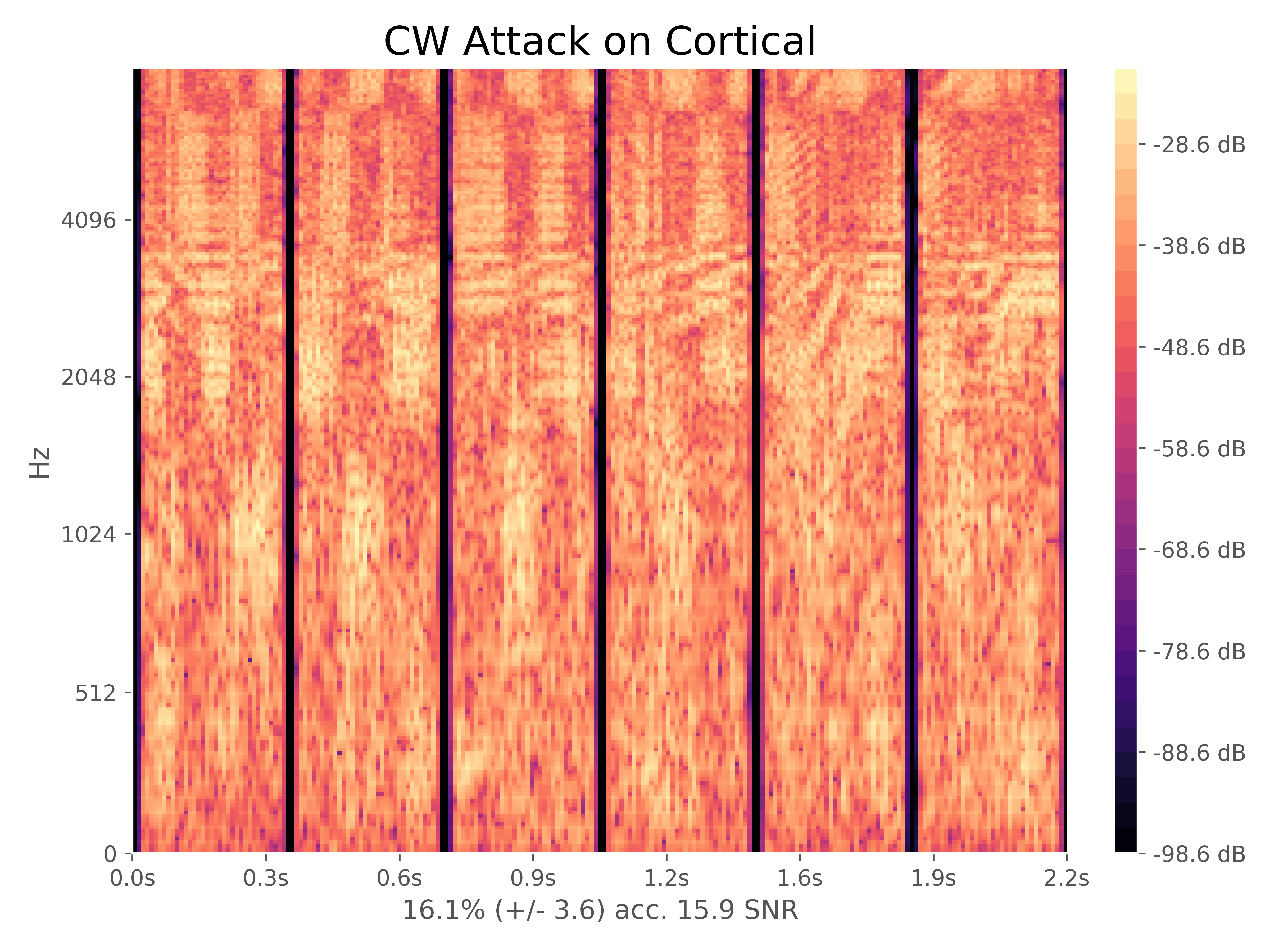}}
	\end{minipage}
	
	\caption{Qualitative differences between the adversarial attacks on baseline and cortical networks for the Deepfool and Carlini-Wagner attacks in the mel spectrogram domain. Each subfigure concatenates the best noise from 6 trials, and reports the SNR and accuracy on these 6 noises on the x-axis. The cortical network attacks appear to pulsate more in time while the baseline network attacks are more stationary. The cortical attacks also exhibit more diagonal lines at resonant frequencies.}
	\label{fig:qual2}
\end{figure}

\subsection{Analysis}

The baseline and cortical networks were both able to train to similar competitive rates as seen in the Detection Error Tradeoff (DET) curves in \cref{fig:perf}.
These curves are specific to the dataset that we created, but for comparison,
the company PicoVoice has created a dataset in a similar manner to ours and tested the CMU Sphinx Open Source Toolkit PocketSphinx, the open source KITT.AI, and their own commercial product PicoVoice.
They advertise a miss rate of 6\% at 1 false alarm per 10 hours for their own product, 32\% for KITT.AI, and 48\% for PocketSphinx.
On our dataset our miss rate is 7\% at 1 false alarm per 10 hours for our baseline and 10\% for the cortical network.
However this is an extreme mode of operation, at 1 false alarm per 1 hour our miss rate is 5\% for both models.

The FGSM and PGD attacks have the slowest accuracy decay rate as the strength of the noise grows on the cortical network.
This is likely due to the more dispersed patterns of attacks that these two methods create.
Because of the STRFs' selectivity for changing phoneme dynamics \cite{mesgarani_2011}, the cortical network should be more invariant to stationary noises as is the human auditory cortex \cite{khalighinejad_2019}, and is only slowly impacted by this type of noise.
We see the noises generated by the Deepfool and Carlini-Wagner attacks are very textured in \cref{fig:qual2}.
For the same search in a given $\ell_\infty$ size ball, the cortical network requires a more elaborate noise than the baseline network off of the auditory spectrogram (and still performs better at the same SNR).
The noise for the baseline is more stationary and has a higher impact on speaking frequencies.
However, all the adversarial noises cannot be said to be imperceptible to the human ear, though speech is still clearly heard and understood by a human, the fricative like quality of the adversarial noises is perceptible even at 25dB SNR.

In the right hand side of \cref{fig:perf} we create adversarial music attacks \cite{adv_music}. These are also defended against by the cortical representation, as seen by the greater decrease in AUC by the baseline. These sounds are much more subtle since they are hidden in guitar like plucks.

\section{Conclusion \& Further Work}

We apply several white-box iterative optimization-based adversarial attacks to an implementation of Amazon Alexa's HW network, and a modified version of this network with an integrated cortical representation, and show that the cortical features help defend against universal adversarial examples.
At the same level of distortion, the adversarial noises found for the cortical network are always less effective for universal attacks.
Further work could start by refining the auditory preprocessing, a more faithful auditory spectrogram would simulate the transduction of the hair cells and the reduction of the lateral inhibitory network.
We also constrained our analysis to highly effective but not imperceptible adversarial noise. Further work could create more subtle noises that degrade performance gradually, enforcing imperceptibility with perceptual masking \cite{qin_2019}.

\vfill\pagebreak

\bibliographystyle{IEEEbib}
\bibliography{refs}

\end{document}